\documentclass[pre,preprint, showpacs,amsmath,amssymb]{revtex4}

\usepackage{graphicx}
\usepackage{epstopdf}
\usepackage{dcolumn}
\usepackage{bm}
\usepackage{multirow}
\usepackage{natbib}

\begin{document}
\preprint{Makita}

\title{Fast electron propagation in Ti foils irradiated with sub-picosecond laser pulses at $I\lambda^{2}  > 10^{18}$ Wcm$^{-2} \mu m^{2}$.} 



\author{M Makita}
\affiliation{Centre for Plasma Physics, School of Mathematics and Physics, Queen's University Belfast, University Road, Belfast BT7 1NN, UK}
\author{G Nersisyan}
\affiliation{Centre for Plasma Physics, School of Mathematics and Physics, Queen's University Belfast, University Road, Belfast BT7 1NN, UK}
\author{K McKeever}
\affiliation{Centre for Plasma Physics, School of Mathematics and Physics, Queen's University Belfast, University Road, Belfast BT7 1NN, UK}
\author{T Dzelzainis}
\affiliation{Centre for Plasma Physics, School of Mathematics and Physics, Queen's University Belfast, University Road, Belfast BT7 1NN, UK}
\author{S White}
\affiliation{Centre for Plasma Physics, School of Mathematics and Physics, Queen's University Belfast, University Road, Belfast BT7 1NN, UK}
\author{B Kettle}
\affiliation{Centre for Plasma Physics, School of Mathematics and Physics, Queen's University Belfast, University Road, Belfast BT7 1NN, UK}
\author{B Dromey}
\affiliation{Centre for Plasma Physics, School of Mathematics and Physics, Queen's University Belfast, University Road, Belfast BT7 1NN, UK}
\author{D Doria}
\affiliation{Centre for Plasma Physics, School of Mathematics and Physics, Queen's University Belfast, University Road, Belfast BT7 1NN, UK}
\author{M Zepf}
\affiliation{Centre for Plasma Physics, School of Mathematics and Physics, Queen's University Belfast, University Road, Belfast BT7 1NN, UK}
\author{CLS Lewis}
\affiliation{Centre for Plasma Physics, School of Mathematics and Physics, Queen's University Belfast, University Road, Belfast BT7 1NN, UK}
\author{A.P.L. Robinson}
\affiliation{Central Laser Facility, Rutherford-Appleton Laboratory, Chilton, Didcot, OX11 OQX, UK}
\author{S.B. Hansen}
\affiliation{Sandia National Laboratory, Albuquerque, New Mexico 87123,USA}
\author{D Riley}
\email{d.riley@qub.ac.uk}
\affiliation{Centre for Plasma Physics, School of Mathematics and Physics, Queen's University Belfast, University Road, Belfast BT7 1NN, UK}


\date{\today}

\begin{abstract}
We have studied the propagation of fast electrons through laser irradiated Ti foils by monitoring the emission of hard X-rays and K-$\alpha$ radiation from bare foils and foils backed by a thick epoxy layer. Key observations include strong refluxing of electrons and divergence of the electron beam in the foil with evidence of magnetic field collimation. Our diagnostics have allowed us to estimate the fast electron temperature and fraction of laser energy converted to fast electrons. We have observed clear differences between the fast electron temperatures observed with bare and epoxy backed targets which may be due to the effects of refluxing.\end{abstract}


\maketitle 

\section{Introduction}
In high intensity laser-plasma interactions it is common to generate a population of so called fast (or hot) electrons, with typical energies ranging from several tens of keV to several MeV \citep{rousse,reich,eder,wharton,nishimura,kmetec}. The dynamics of such fast electrons are important in fast ignition fusion. More relevant to this work, they are also influential in determining the size, duration and efficiency of short pulse X-ray sources used in many scientific experiments as a probe and as a diagnostic  \citep{ park,serbanescu,chen,nishikino}. 

In this paper we have used observations of K-$\alpha$ emission and hard x-rays to infer information about the dynamics of the fast electrons generated when a sub-picosecond pulse at 1.053 $\mu$m wavelength is incident on a Ti foil at peak intensity $>10^{18}$ Wcm$^{-2}$. We have measured absolute yields of K-$\alpha$ photons and hard x-rays. Using the hard x-ray emission, we have inferred an effective temperature for the fast electrons as well as total conversion into hard x-rays generated by them. The expansion and penetration of fast electrons through the foil is monitored by imaging the K-$\alpha$ signal from the rear of the foils as a function of thickness. The penetration of the electrons into the foil and the contribution of refluxing of electrons to the K-$\alpha$ signal are investigated by observing the emission of K-$\alpha$ radiation from targets with and without a thick (~1 mm) layer of epoxy on the rear. In the following sections we shall first describe the experimental geometry, targets and instruments. We then move on to presenting the data, comparison with modelling and our conclusions.

\section{Experiment}
\subsection{Diagnostics}
The experiment was carried out with the high power laser system, TARANIS \citep{dzelzainis} situated at Queen's University Belfast. This Nd:Glass chirped-pulse-amplified laser can provide pulses of 800 fs full-width at half-maximum (FWHM) duration at 1.053$\mu$m wavelength. The ASE intensity contrast of the laser at 2 ns before the main pulse was measured to be $10^{-7}$. The pre-pulse activity consisted of a few, picosecond duration, pre-pulses at up to approximately 2.4ns ahead of the main pulse with intensity contrast of $2\times10^{-7}$ compared with the main pulse.

The p-polarized beam was focused by an F/3.3 off-axis parabola (OAP) to a focal spot of 12$\mu$m full-width-at-half-maximum (FWHM) diameter containing 50$\%$ of the energy. The size of the focal spot was inferred from preliminary shots at full energy in which a 4.6m focal length lens was used to focus the full beam down onto a 12 bit CCD camera after reflection off several high quality optical flats to reduce energy to a level where appropriate filtering could be used to make measurements. In this experiment we used energies up to 5J. Defining the peak intensity as $I =(E/\pi r^{2}t_{p})$cos$\theta$ where E is the laser energy in the central spot, t$_{p}$ is the FWHM pulse duration, $\theta$ is the angle of incidence way from normal and 2r is the FWHM diameter of the central spot, we estimate a peak intensity of about $2\times10^{18}$ Wcm$^{-2}$ for 40$^{\circ}$  incidence on target for a 5J shot.

\begin{figure}[htbp]
\centering
  \includegraphics[width=7.5cm]{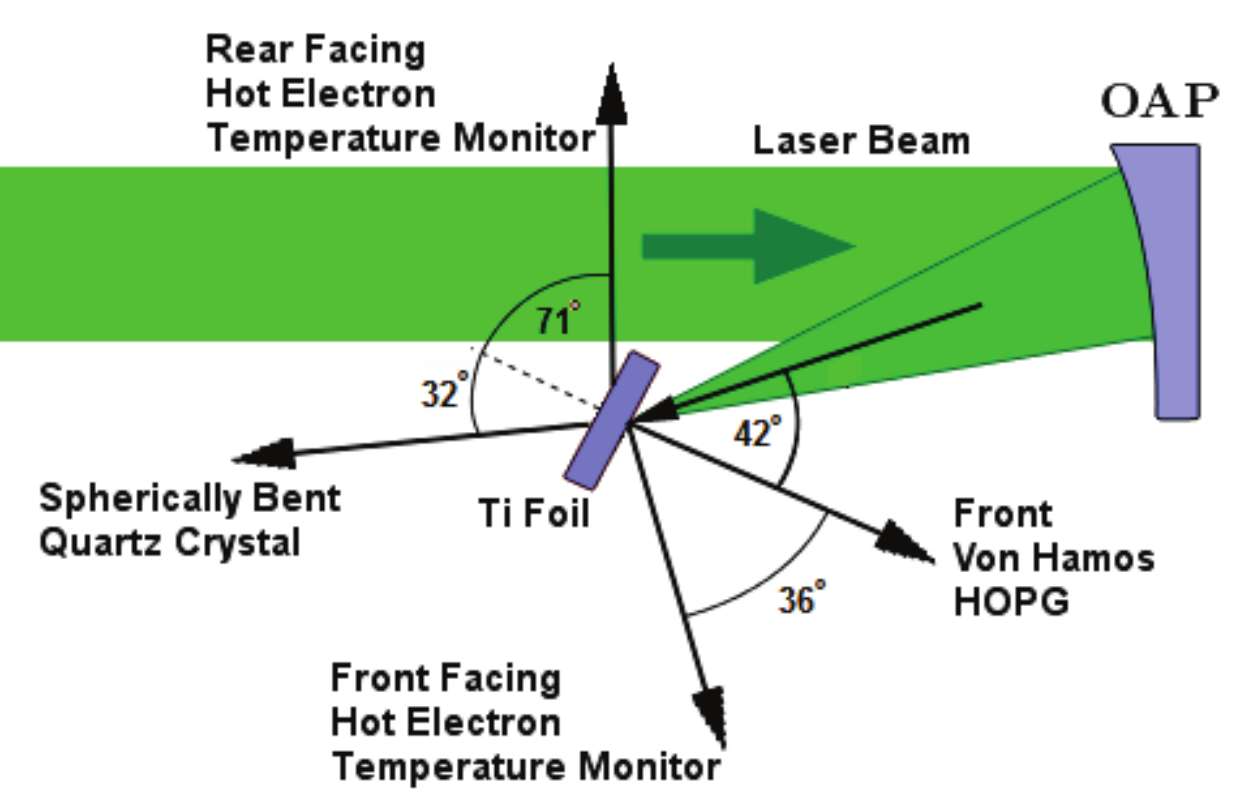} \\[2pt]
  \caption{(Color online)Schematic of experimental arrangement. In this figure, the laser incidence is $40^{\circ}$ to normal and the front  HOPG spectrometer views 2$^{\circ}$ away from the front target normal. The spherical crystal spectrometer viewed 32$^{\circ}$ from the rear target normal. The front facing hot electron temperature monitor position was not used in this experimental run.}
  \label{layout}
\end{figure}

The K-shell emission from laser-irradiated thin foils of Ti was monitored with a spectrometer observing at near normal to the front, laser irradiated side, of the foil. The spectrometer was operated in the von-Hamos configuration using a cylindrically curved (R$=50mm$) highly oriented pyrolytic graphite (HOPG) crystal ($2d=6.708$ \mbox{\AA}) coupled to an image plate (IP) detector. The layout is shown schematically in figure \ref{layout}. The spectral range covered was sufficient to include the K-$\alpha$ and K-$\beta$ lines (2.75  \mbox{\AA} and 2.51\mbox{\AA} respectively) as well as the He-like resonance line at $2.62$  \mbox{\AA} and its Li-like satellites as well as the H-like resonance line and satellites at $2.49$ \mbox{\AA}. The integrated reflectivity of the HOPG crystal was determined to be $1.9\times10^{-3} $rad$ \pm15\%$ at 2.75  \mbox{\AA} by comparison of spectra with a single hit CCD detection system using low laser energy shots in a separate experiment. 

A spherically bent quartz crystal (R$=150mm$, $2d=3.082$ \mbox{\AA}) was placed to view the rear of the target foils (the side away from laser irradiation). This instrument provided high spectral resolution in the region around the K-$\alpha$ emission as well as one-dimensional spatial resolution with a magnification of 1.1.  This allowed us to observe changes in the K-$\alpha$ line shape caused by heating of the bulk foil and also to observe the size of the emitting region in order to measure the expansion of the electrons in the foil. Since electro-magnetic pulse damage obliged us to use an image plate rather than a CCD, the spatial resolution was limited by the resolution of the image plate. 

By using a series of shots on a $50 \mu$m Ti wire, we established that the spatial resolution of the spherical crystal instrument in the target plane is about $70\mu$m. Previous work with a similar crystal and experiment suggests we should expect a spectral resolving power of $\lambda/\Delta\lambda \sim3800$ \citep{hansen1}; folding in the resolution of the image plate, this is reduced to ${\lambda}/{\Delta\lambda} \sim1600$.

In order to estimate the fast electron temperature, we fielded a simple instrument that measured the bremsstrahlung radiation generated by interaction of the fast electrons with the foil. This consisted of an array of six filters (with thicknesses ranging from 50-300 $\mu$m of Pb) with an image plate as the detector. The principle of the instrument is based on the assumption that the bremsstrahlung emission has a spectral shape given by $I(E_{\nu},T_{h}) \sim exp(-E_{\nu}/k_{B}T_{h})$ where $T_{h}$ is the fast electron temperature, $k_{B}$ is the Boltzmann constant and $E_{\nu}$ is the photon energy. 

A model that predicts the relative signals through the filters by folding in the image plate response to hard x-rays (Fuji MS type)  \citep{doppner,meadowcroft} and transmission through lead filters \citep{nist} was used to generate the effective fast electron temperature. The image plate has been calibrated out to 662keV \citep{doppner}.  The basic assumption that we have an exponential slope for bremsstrahlung determined by $T_{h}$ has been discussed by McCall \citep{mccall} who has pointed out that it is valid as long as our measurements are dominated by photons with energy $E_{\nu} > k_{B}T_{h}$. To test the validity of our diagnostic, we have used a simple computer model based on the empirical observation that for a single electron energy, $E_{e}$, impinging on a foil, the bremsstrahlung energy emission spectrum scales as $I = CZ(E_{e}-E_{\nu})^{\alpha}$ where $E_{\nu}$  is the photon energy, $\alpha$ is a constant close to unity (1.15 for Ti) and Z is the atomic number of the foil material; see for example \citep{mccall,storm}. With this, we have predicted the signal that would be detected through our filter array. We have determined that for a purely exponential electron distribution (2-dimensional Maxwellian), the assumption of an exponential spectrum in analysing the bremsstrahlung is accurate to within the statistical error bars on our data. However, if the fast electron distribution is 1-dimensional then there is a systematic underestimate of temperature by up to 20$\%$ at the highest values and an overestimate of similar magnitude if the electron distribution was fully 3-dimensional. The data presented here assumes a 2-dimensional Maxwellian. An alternative approach is to use the Bethe-Heitler cross-section corrected for lower energy electrons as discussed by Salvat {\it et al} \citep{salvat}. Using this, we get agreement to within about 5$\%$ with the empirical approach discussed above. We prefer the latter in analysing our data as it is based on experimental observation and is faster computationally.

The viewing angle in the horizontal plane was at 71$^{\circ}$ to the rear target normal and 35$^{\circ}$ above the plane of the interaction. The filter array and IP sat outside the chamber and a 50 $\mu$m mylar window allowed X-rays to pass. A collimating tube covered in lead (thickness 1mm) and with a slot at the front ensured that the instrument viewed only X-rays from the region of the target and helped to reduce the plasma striking the window and causing fluorescence. The window also served to slow fast electrons (stopping all below 60 keV) reaching the Pb filters. In addition, a pair of magnets placed between the instrument and the target generated an 0.1T magnetic field to deflect fast electrons from the instrument. Some fluorescence from the filter array was observed on the image plate, and was monitored in preliminary shots as a function of distance between filters and image plates and was removed in analysis. For the range of temperatures we observe below, the data is reliable generally for 5 filters; the thickest showing little signal. 
	
\subsection{Targets}
In this work we have used Ti foils of approximately 5 mm $\times$ 5 mm with several mounted on a frame holder at a time. The foils were of 10, 25, 50, 75 and 100 $\mu$m thickness. In some data runs a thick ($\sim$1mm) layer of epoxy (A/epichlorohydrin,C$_{21}$H$_{25}$ClO$_{5}$) was bonded on the rear side of the foils. The purpose was to allow us to explore the effect of refluxing by comparing K-$\alpha$ and hard x-ray yield as a function both of target thickness and the presence of the epoxy layer. We shall see below that the typical fast electron energy is of order 80 keV. Collisional stopping powers \citep{nist2} indicate a range of only about 50 $\mu$m in the epoxy. For a 1mm layer, we expect only electrons with energy in excess of 800 keV to make it to the rear and return to the Ti. Furthermore, in a dielectric a large electric field can develop that inhibits fast electron transport. This has been discussed by Quinn {\it et al} \citep{quinn} following Tikhonchuk \citep{tikhonchuk}. In the latter work, it is explained that for a dielectric target, a large inhibiting electric field can be generated by fast electron penetration. This field can be large enough to cause significant ionisation of the dielectric and this is a mechanism for dissipation of energy from the electron beam that can exceed the collisional loss rate. The loss rate is dependent on several factors, including the density of fast electrons, the average ionisation potential of the dielectric and the Coulomb logarithm; values of 1-10keV/ $\mu$m were estimated\citep{quinn,tikhonchuk}.  Thus it is supposed, in this work, that fast electrons that reach the rear of the foil are stopped in the epoxy and prevented from returning (refluxing) to create more inner shell emission, an approach taken in previous similar experiments under different conditions \citep{nersisyan,neumayer}. The purpose of the spatial/spectral imaging with the spherical crystal is to observe any divergence of the electron beam as it passes through the foil and to observe bulk heating effects from changes in the spectrum of K-$\alpha$  \citep{hansen2}. We have looked at the relative K-$\alpha$ yield as a function of angle of incidence of the laser on target to help decide the absorption mechanism. In addition, the hard x-ray measurements have helped us to establish a fast electron temperature and also helped to determine the dominant absorption mechanism. In the next section, we present data obtained and discuss its interpretation using simulations of varying degrees of sophistication. 

\section{Experimental Data}
\begin{figure}[htbp]
\centering
  \includegraphics[width=7.5cm]{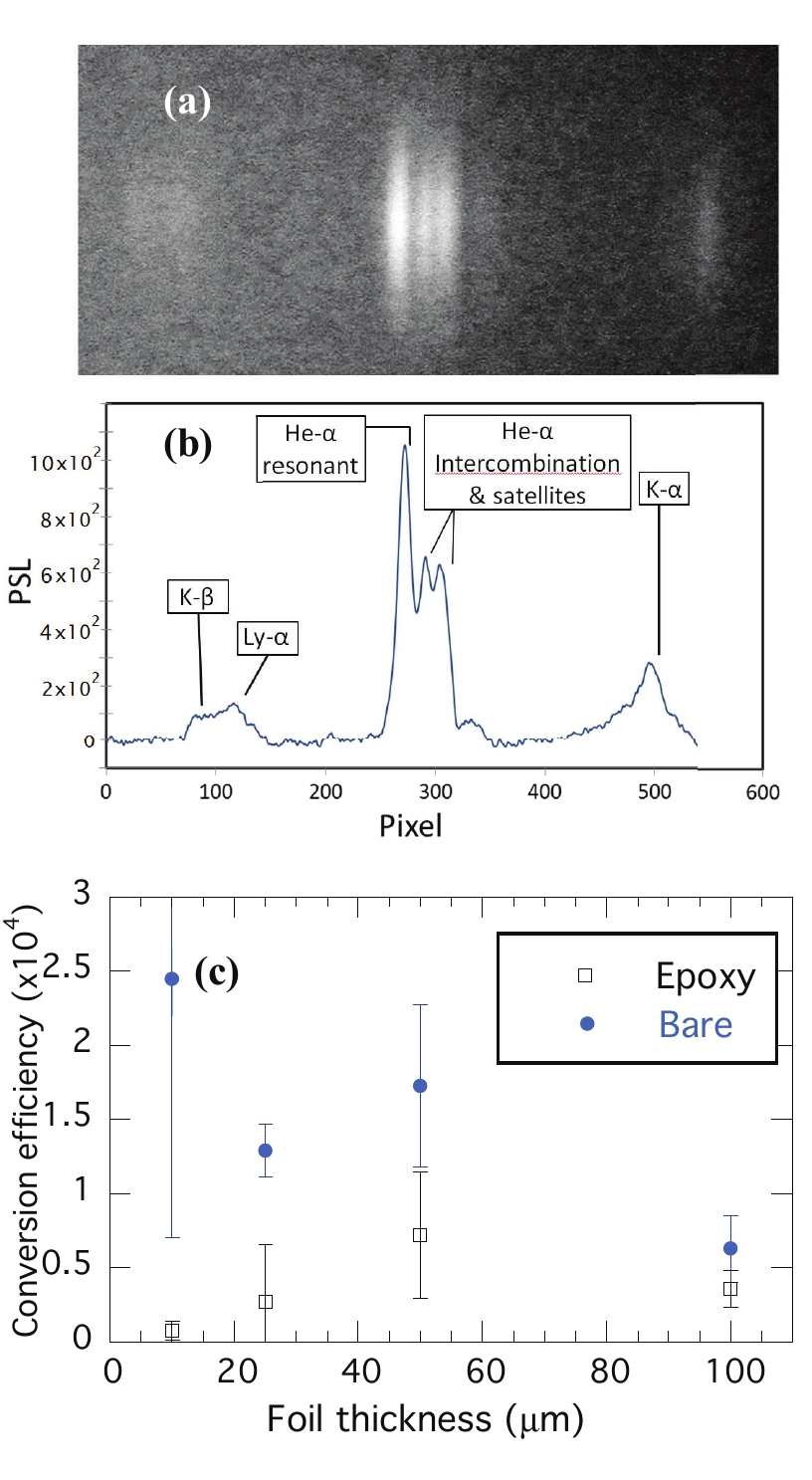} \\[2pt]
  \caption{(Color online) (a) Typical spectrum from the HOPG crystal looking at the front of a 50$\mu$m Ti foil with the laser incident with 5J at 40$^{\circ}$ to target normal. (b) averaged line-out of data (c) variation in He-$\alpha$ yield as a function of foil thickness and presence of epoxy backing.}\label{spectrum}
\end{figure}

In figure \ref{spectrum} we see a typical spectrum from the HOPG crystal for angle of incidence 40$^{\circ}$. We can see the Ti K-$\alpha$ line as well as the He-$\alpha$  group (He-like 1s$^{2}$-1s2p $^{1}$P, 1s$^{2}$-1s2p $^{3}$P and Li-like satellites) and some weaker H-like emission. The FLYCHK code \citep{flychk} has been used to find the best fit to the ratio of resonance line, inter-combination line, Li-like satellites and the Ly-$\alpha$ line. We deduce that the thermal electron temperature of the pre-formed plasma is 0.8$\pm$ 0.05 keV at the critical density of N$_{e}$ =10$^{21} cm^{-3}$. We have assumed, in the simulation, a hot electron temperature of 60 keV with a hot electron fraction of 10$^{-3}$ as this combination gave the best fit to the spectrum. We should note that the rapid timescale of heating may make time dependent effects important and this temperature is an estimate based on spectra that are time and space integrated. An interesting phenomenon can be seen in figure \ref{spectrum}(c). The He-$\alpha$  emission has wide variation from shot to shot, as evidenced by the large standard deviation in the mean of several shots shown in the figure, but it is clearly affected by the presence of the epoxy layer, especially for the thinnest foils. The signal is too weak for the epoxy case to determine a good fit to temperature from FLYCHK and the shot to shot variation is higher than for the K-$\alpha$ radiation. However, we can deduce that the refluxing fast electrons are important in determining the spectrum emitted by enhancing excitation of the He-like ions, either by exciting collisions or providing additional heating in the pre-formed plasma.

\begin{figure}[htbp]
\centering
  \includegraphics[width=7.5cm]{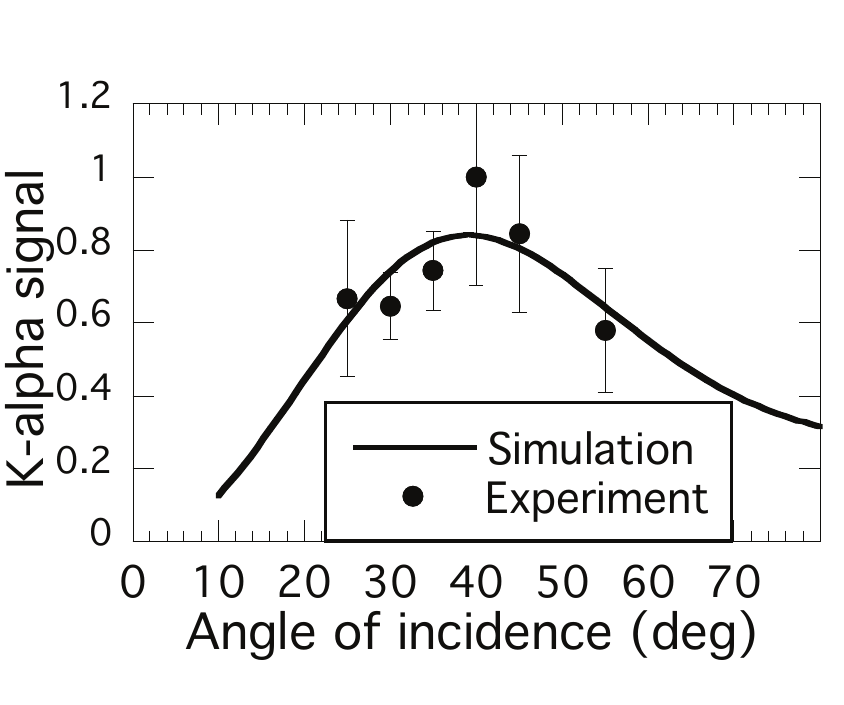} \\[2pt]
  \caption{Angular K-$\alpha$ variation, for 50$\mu$m Ti foils, compared to a resonance absorption model assuming that K-$\alpha$ emission is proportional to absorption.}\label{anglescan}
\end{figure}

In figure \ref{anglescan}, we see the relative K-$\alpha$   signal as a function of angle of incidence for 50 $\mu$m foils. We have plotted in normalised units to compare with a simple resonance absorption model \citep{kruer}, although the peak corresponds to a conversion efficiency from laser energy to K-$\alpha$ photon energy of approximately 10$^{-4}$. There is an optimum angle of around 40$^{\circ}$. However, the error bars due to shot-to-shot variation make this uncertain. If we assume, for the present, that resonance absorption is the dominant mechanism, we can make an estimate of the plasma scale-length. In doing this we make a relatively straightforward assumption that the K-$\alpha$   emission is only weakly sensitive to intensity (which varies with angle) and is linearly dependent on absorption. The result of a least squares best fit process is plotted with the experimental data in figure \ref{anglescan}, and we have a scale-length, L=0.34 $\mu$m and thus L/$\lambda \sim $0.3.

Hydrodynamic simulations of the pre-pulse interaction with the HYADES code \citep{larsen} predict a pre-formed plasma with scale-length $ L=2.4 \mu$m and a temperature of $\sim 50 $eV in the under-dense region. Thus, ponderomotive steepening of the plasma is indicated and this is perfectly reasonable for the intensities used. For our peak intensity, we have an electron excursion length of $v_{osc}/{\omega} \sim0.2 \mu$m This regime is appropriate for the resonance absorption mechanism. However, due to the shot to shot variation, the fit to resonance absorption is not perfectly clear cut and we cannot discount other mechanisms such as the ${\textbf J \times \textbf B}$ acceleration mechanism on the basis of this data. This latter mechanism is applicable in the relativistic regime that we are in since the normalised vector potential a$_{0} \sim 1$ where;

\begin{eqnarray}
a_{0}= \sqrt{    \frac{I\lambda^{2}}{1.37\times10^{18} Wcm^{-2} \mu m^{2}}  }
\end{eqnarray}

This mechanism is characterised by acceleration of bunches of electrons at twice the laser frequency and strong 2$\omega_{L}$ emission from the rear of the foil as they exit the foil. Such a diagnostic was not available on this experiment. However, we can turn to the front-side hard x-ray emission diagnostic to help resolve the issue.

\begin{figure}[htbp]
\centering
  \includegraphics[width=7.5cm]{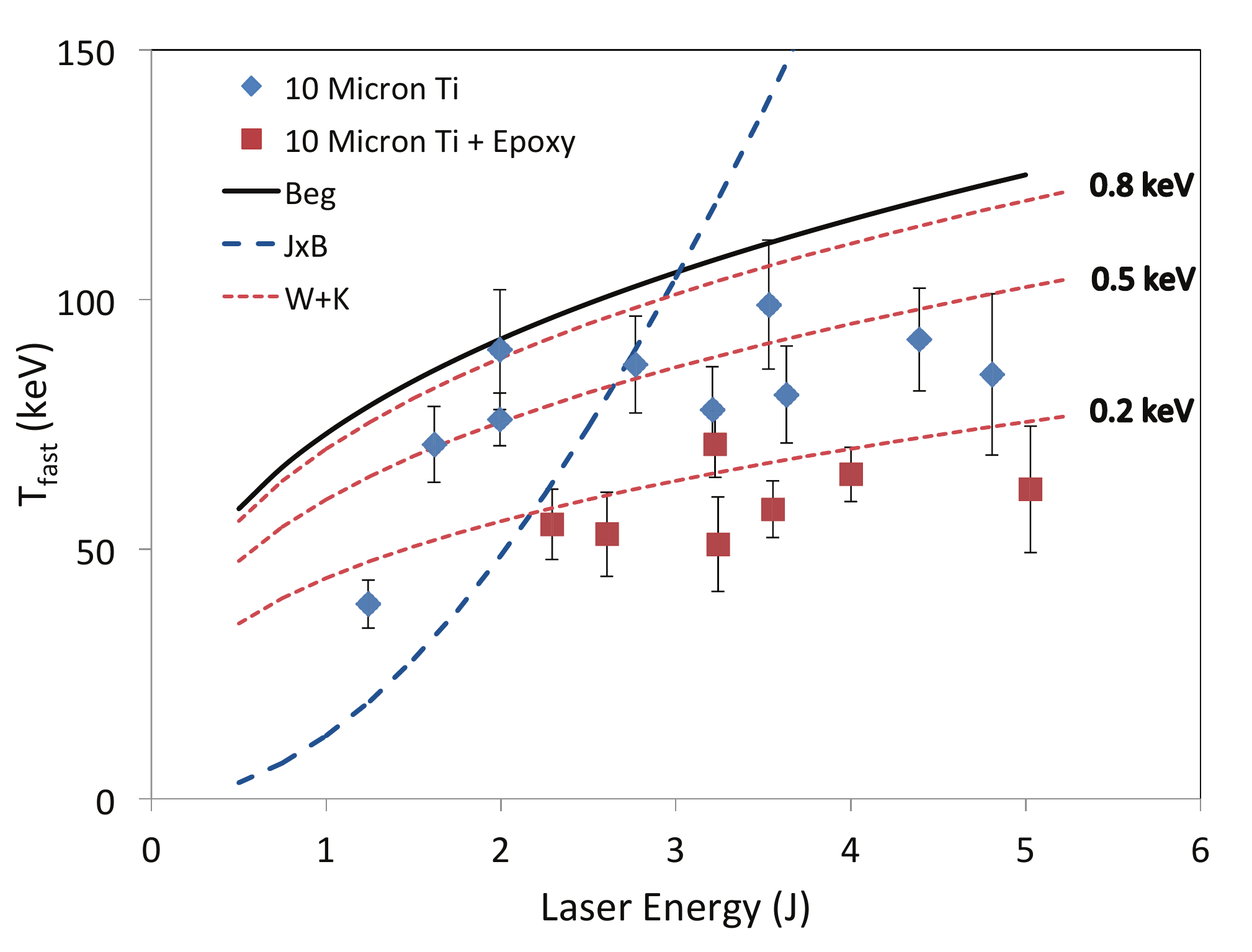} \\[2pt]
  \caption{(Color online) Inferred hot electron temperature from the hard x-ray emission, assuming a purely exponential electron distribution. All shots at 40$^{\circ}$ incidence. There is a clear difference between targets with and without an epoxy backing.The curves representing different fast electron temperature scalings are discussed in the text.}\label{hottemp}
\end{figure}

In figure \ref{hottemp}, we can see the inferred fast electron temperatures for shots on 10 $\mu$m thick foils, with and without epoxy backing. The temperatures are mostly between 50-100 keV for bare foils. As discussed above, the electrons are assumed to have a single effective Maxwellian distribution and the assumed bremsstrahlung spectrum is convolved with the filter transmissions and image plate response to find a best fit to the ratios of signals in the different filter channels. We used all possible pairs of filters to obtain temperature estimates for each shot. The variance in the mean of these provides the error bars. For epoxy backed foils there is a distinct drop in inferred temperature. One intuitive interpretation is that this is due to a preferential loss of fast electrons into the epoxy. We tested this hypothesis by constructing a computer model for bremsstrahlung emission based on a corrected Bethe-Heitler bremsstrahlung cross-section \citep{salvat}, as discussed above, with stopping power for electrons from NIST \citep{nist2}. We compared emission spectra from cases where we assume a thin (10$\mu$m) foil and allow electrons to traverse the foil once with simulated spectra where all electron groups up to initial energy 10kT are allowed to reflux until they dissipate all energy.  This model did indeed indicate such an effect, however, when we fitted the resultant spectra to an exponential to get effective temperatures, the difference between the target types was only of order 3-4 keV, well inside the error bars.  This means that preferential loss of faster electrons into the epoxy seems unable to explain the experimental difference. One reason for this is connected to the fact that for the detection system, the filter transmission falls rapidly below about 40keV and only electrons above this energy are expected to reach the rear of the foil. This means that the electrons responsible for the measured signal are generally either all absorbed in the epoxy after the first pass through the foil or all reflux after one pass, depending on target type. Thus the emission drops, as seen in figure \ref{hardxray} below, but there is little change in the effective spectral shape. Additionally, as discussed below, the epoxy makes a finite contribution to the bremsstrahlung emission and thus the difference is expected to be even smaller. We discuss an alternative explanation below. For comparison, we have plotted some fast electron scaling laws available from the literature.  The solid line represents the experimental scaling of Beg $\textit{et al}$ \citep{beg}, T$_{h} \sim 100(I/10^{18}Wcm^{-2})^{1/3}$. We see that it overestimates the temperature in our case. The thicker dashed line is the scaling expected for the ${\textbf J \times \textbf B}$ mechanism for p-polarization \citep{mtv};

\begin{eqnarray}
T_{h}=m_{0} c^{2} [(1+a_{0}^{2}/2)^{1/2}-1]
\end{eqnarray}
For resonance absorption, we can turn to Wilks and Kruer \citep{wilkskruer} for an estimate of expected temperature in keV;

\begin{eqnarray}
T_{h} \approx 10[T_{b} I_{15} \lambda^{2} ]^{1/3}   
\end{eqnarray}
where T${_b}$ is the background electron temperature in keV, I$_{15}$ is the intensity in units of 10$^{15}$ Wcm$^{-2}$ and $\lambda$ is the wavelength in $\mu$m . Curves for Wilks and Kruer (thinner dashed lines) are shown in figure \ref{hottemp} with different values of  T${_b}$. In practice, the background temperature would also vary with incident energy but we use curves of constant background to keep the discussion simple. Assuming a background temperature of about 1 keV would give us the Beg scaling law.

We estimated, earlier, from spectral data, for a typical 4J shot, that the background plasma temperature T${_b} \sim 0.8$ keV at critical density. For our bremsstrahlung data, a more modest background temperature of around 0.5 keV seems to fit better and the scaling with laser energy seems acceptable. As noted above, the spectrometer is a time and space integrated diagnostic. Also, the peak fast electron generation is at the peak of the pulse when only half the laser energy has been delivered. Thus a background temperature of closer to 0.5 keV is  consistent with our bremsstrahlung data. In any case, it seems that a scaling law more compatible with resonance absorption than ${\textbf J \times \textbf B}$ acceleration is more appropriate to explain our data. It is worth noting that the comparisons made in figure \ref{hottemp}, are largely between the temperature generated "at source" by the two mechanisms considered and the temperature inferred from spatially and temporally averaged bremsstrahlung emission. The Beg scaling is of course derived experimentally and thus includes effects of electron transport as well as spatial and temporal averaging.

The dependence of the fast electron temperature on the background plasma temperature suggests another possible reason for the difference in effective temperature between bare and epoxy coated foils. For the average electron energy, the time for an electron to transit a 10$\mu$m foil and back is only of order 0.13ps, well below the pulse duration. Thus we might consider the possibility that fast electrons generated on the rising edge of the main pulse may return to alter the front plasma conditions before the peak of the pulse, thus altering the time and space integrated effective fast electron temperature. 
\begin{figure}[htbp]
\centering
  \includegraphics[width=7.5cm]{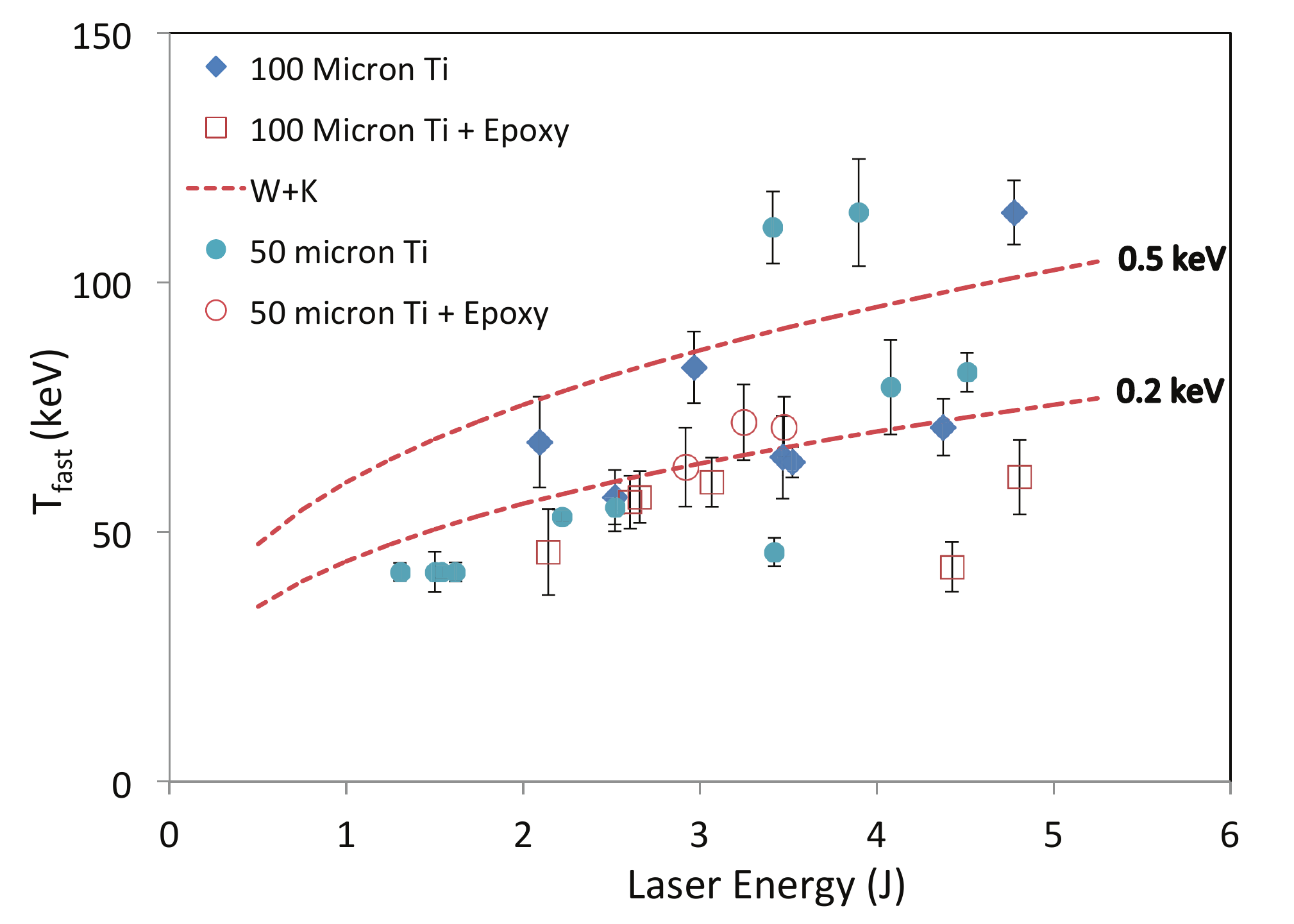} \\[2pt]
  \caption{(Color online) Inferred hot electron temperature from the hard x-ray emission for thicker foils with and without epoxy backing.}\label{thickvsepoxy}
\end{figure}

However, we can predict that this effect would not be noticed for the thickest foils due to the longer transit time and loss of electron energy in the foil. In figure \ref{thickvsepoxy} we show data for 50 and 100$\mu$m foils comparing bare and epoxy coated targets. There are some data points for bare targets with temperatures clearly above coated foils but, overall, the difference is far less well defined than for the thin foil data. Additionally, we can note that, for bare foils, the typical inferred fast electron temperature is lower than for the thin foil case in figure \ref{hottemp}. This would also be an expected consequence of the main laser-plasma interaction being affected by refluxing from thin foils.

We can make an estimate of the likely level of heating caused by refluxing electrons entering the pre-formed plasma with some simple assumptions. As we shall see below, further data will allow us to make estimates of the number of fast electrons that reflux from the rear of the foil ($\sim10^{14}$) and their average energy when they enter the pre-formed plasma ($\sim$100 keV)  as well as the area over which fast electrons are injected into the target  (70$\mu$m diameter), which is set to the minimum K-$\alpha$ source size discussed below. We then use the stopping power calculations of Li and Petrasso \citep{li} to estimate the average energy loss rate of fast electrons in the plasma at around critical density ( $\sim$ 6 MeV cm$^{2}$/g). This leads however to an estimate that the upper limit is approximately 30 eV additional heating, which seems insufficient for this explanation to work. Thus, although it seems reasonable to conclude from the data that refluxing is connected to the effective temperature measured for the fast electrons, it is unclear at this point what the physics of the responsible mechanism is.

\begin{figure}[htbp]
\centering
  \includegraphics[width=7.5cm]{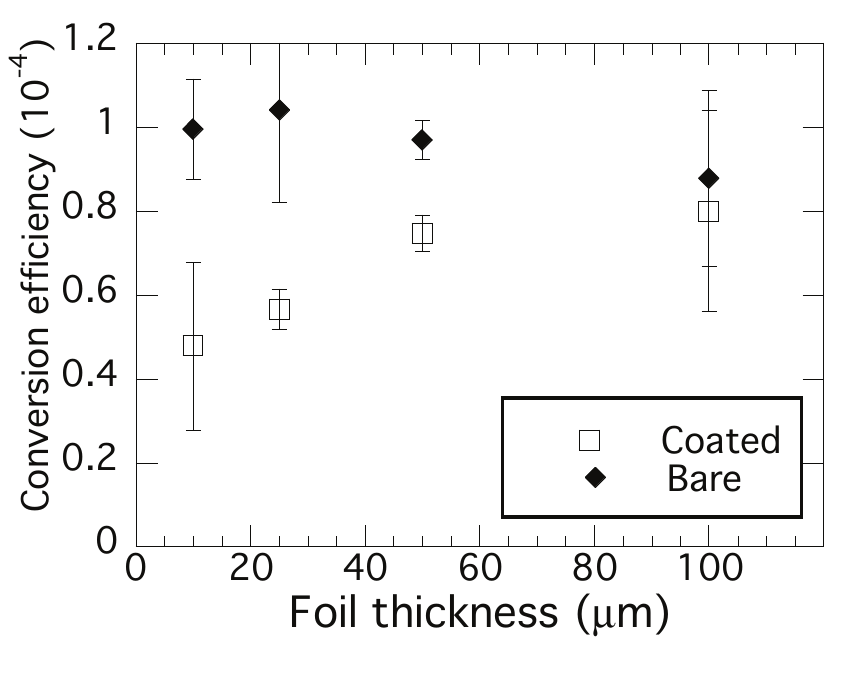} \\[2pt]
  \caption{K-$\alpha$ yield, viewed from target front side, as a function of foil thickness for both bare and epoxy backed targets. Data points are an average of several shots of 3-5 shots and error bars represent standard deviation in the mean.}\label{kalpha}
\end{figure}

Because we know the geometry of the experiment, the reflectivity for the crystal and the efficiency of the image plates, we can use the measured fast electron temperatures along with absolute K-$\alpha$ emission, in order to get an estimate of absolute absorption into fast electrons. We do this below. 

First, let us consider figure \ref{kalpha}  which shows the measured K-$\alpha$ emission, measured on the HOPG spectrometer, as a function of foil thickness for both bare foils and foils backed by a layer of epoxy. The refluxing of electrons provides about half the total of K-$\alpha$  emission for thin foils. For bare foils, the total emission drops slightly with thickness. This is expected since faster electrons that would reach the back of the thin foil and reflux back to create more signal, instead penetrate deeply and any emission is partially absorbed before it can escape to be detected. By contrast, for the epoxy backed foils, the emission increases with thickness of foil since we have more material contributing to the emission before electrons are 'lost' to the epoxy layer.

\begin{figure}[htbp]
\centering
  \includegraphics[width=7.5cm]{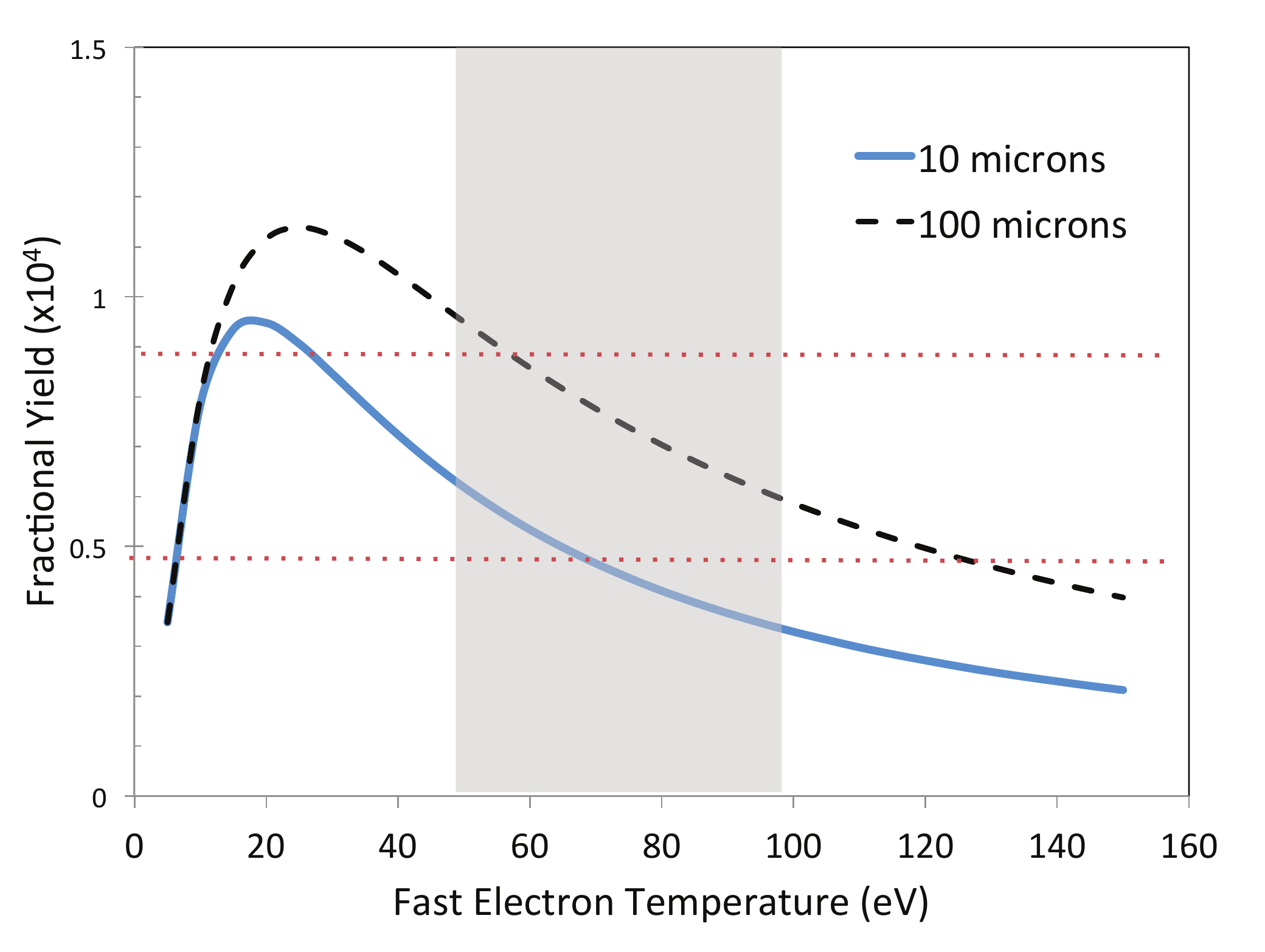} \\[2pt]
  \caption{(Color online) Simulated yields for K-$\alpha$ assuming overall $15\%$ conversion of laser energy to fast electrons for epoxy backed foils (no refluxing). The shaded area indicates the experimental range of hot electron temperatures. The upper dotted horizontal line shows the measured average yield for 100 $\mu$m foils and the lower horizontal line shows measured yield for 10 $\mu$m foils.}\label{simulate1}
\end{figure}

We note that the total fractional yield is of order $10^{-4}$ of the laser energy, in agreement with previous work e.g. \citep{rousse,reich,eder,wharton,nishimura,kmetec}. For the epoxy backed cases, we have assumed that all refluxing is prevented and have simulated the yield using a simple model of K-$\alpha$ generation \citep{khattak}, based on earlier work by Reich {\textit et al} \citep{reich}. We see the results for the thinnest and thickest foils in figure \ref{simulate1}. We note that for an assumption of $15\%$ absorption the yield is just a little lower than the measured value for temperatures in the middle of the measured range. Given that about half the laser energy is in the central focal spot, it is possible to conclude that the conversion to fast electrons is in the range 30-35$\%$ for the central spot. However, we need to exercise caution as the analysis of the optical focus suggests the other half of the laser energy is in an approximately 50 $\mu$m "halo", with an irradiance of $\sim10^{17} Wcm^{-2}$ which, based on an assumed $I^{1/3}$ scaling, is high enough to generate fast electrons with $20-40$ keV energy, although it is not clear what the absorption fraction is likely to be in the "halo" region.

A second way to estimate the conversion into fast electrons is to estimate the absolute hard X-ray yield from the lead filter array data.  We can see in figure \ref{hardxray} measurements of total bremsstrahlung yield as a function of laser energy. This is estimated using the measured fast electron temperatures and assuming the emission follows an exponential distribution, which, as discussed above, should be accurate for a 2-dimensional Maxwellian electron distribution. We are also assuming that the bremsstrahlung from the lower temperature "thermal" plasma, created at the focal spot, contributes little to the measured signal through the lead filters. Following McCall  \citep{mccall} the efficiency of bremsstrahlung emission for our case can be written; $\eta =CZk_{B}T_{h}$ , where, $C=2.2\times10^{-6}$. For a typical fast electron temperature of 80keV we can reproduce our data, with bare foils, for overall conversion of 20-25\% of laser light into fast electrons. This is a little higher than the overall conversion seen in the K-alpha case, but is broadly consistent and we cannot discount that emission from the hot plasma at the focus will contribute to hard x-rays lifting the yield a little above that caused by the fast population alone. For epoxy coated foils, the experimental determination of bremsstrahlung efficiency of the target does not depend on the average Z, but the interpretation of the drop in efficiency compared to bare foils does. The average value of Z$^{2}$ for the epoxy is about twenty times smaller than for the Ti layer. However, we have estimated from collisional stopping that on average the electrons travel through 50 $\mu$m of epoxy. This would, at a crude estimate, mean we expect the epoxy to contribute about a fifth of the total emission seen in a coated target. The contribution of the epoxy means that the drop in signal for coated targets is not so great as for the K-$\alpha$ data.

\begin{figure}[htbp]
\centering
  \includegraphics[width=7.5cm]{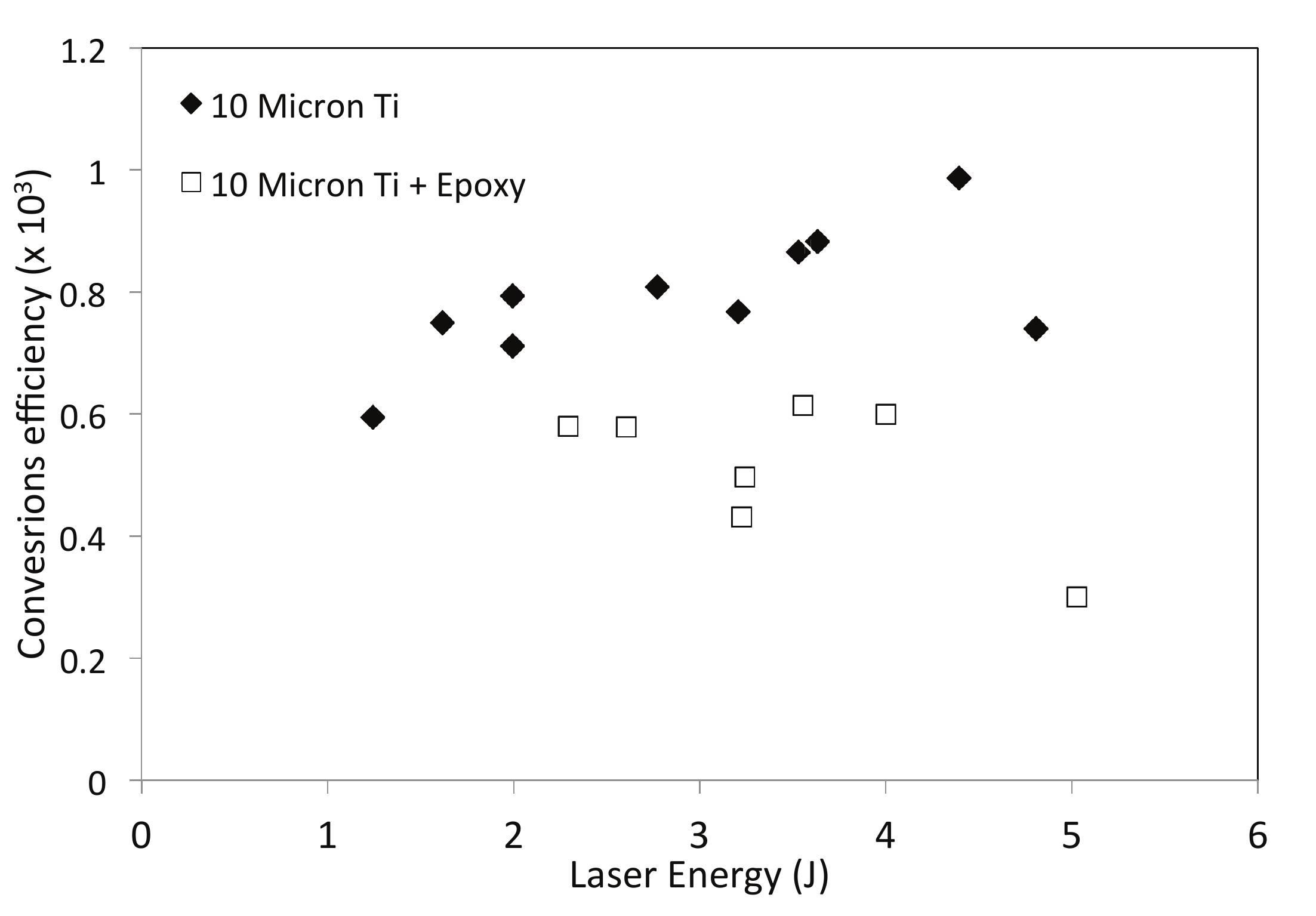} \\[2pt]
  \caption{Conversion efficiency to hard x-rays derived from extrapolation of the lead filter array data.}\label{hardxray}
\end{figure}

\begin{figure}[htbp]
\centering
  \includegraphics[width=7.5cm]{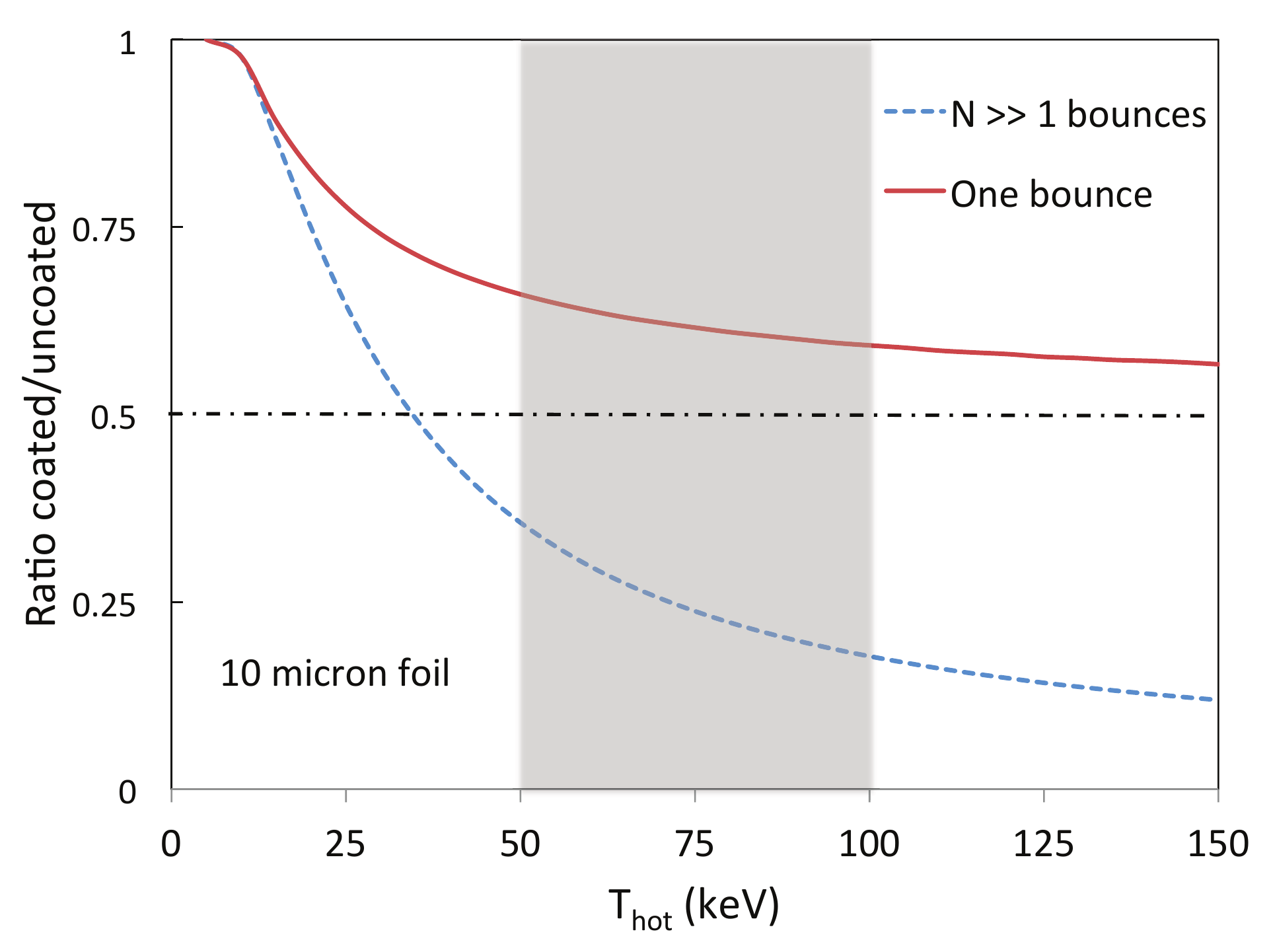} \\[2pt]
  \caption{(Color online) Simulation of the effect of refluxing on K-$\alpha$ emission for 10 $\mu$m foils. N referes to the number of bounces and N$ \gg$1 means that, in the model, each electron group is allowed to reflux from both front and rear surfaces until all energy is lost to collisions. }\label{simulate2}
\end{figure}

Now that we have estimated the numbers of fast electrons and their effective temperature, we will proceed to look at some other elements of the electron dynamics. In figure \ref{simulate2}, we can see the results of 1-dimensional simulations carried out with the simple model that depict the expected ratio of signal between epoxy backed and bare targets for a range of fast electron temperatures (the measured range is shaded). We see two curves; one assumes electrons reflux from front and back of the foil until they lose all energy, the other assumes only one bounce from the rear surface. In this model, the efficiency of refluxing is assumed to be perfect. For our foil size, temperature and energy, this is consistent with the capacitive model results developed by Myatt \textit{ et al} \citep{myatt}, which predicts $>98\%$ efficiency. As above, we assume that the fast electrons penetrate the foil with an exponential distribution. We can see that the experimental ratio of $\sim 0.5$, seen in figure \ref{kalpha}, lies between these curves. This suggests that the effective refluxing is limited, possibly by the effect of the plasma gradient at the front surface but perhaps also by the fact that the electron beam is divergent, as seen in the data below and thus path lengths within the solid are longer. The number of passes through the target in refluxing is likely to depend on initial energy but, to a crude approximation, the experimental ratio between coated and bare targets can be closely reproduced by allowing 2 passes for each group.  We can note that the modelling of Quinn {\it et al} \citep{quinn} indicated many more passes in the refluxing of fast electrons in Cu foils. However, our data does not conflict with theirs since they deal with substantially higher temperatures of order 1 MeV and their model predicts an almost linear relationship between number of passes and initial electron energy. For an average fast electron energy of 80 keV, their modelling predicts  $<$ 2 passes through their 20 $\mu$m Cu foils which is broadly consistent with our conclusions for Ti foils.

\begin{figure}[htbp]
\centering
  \includegraphics[width=7.5cm]{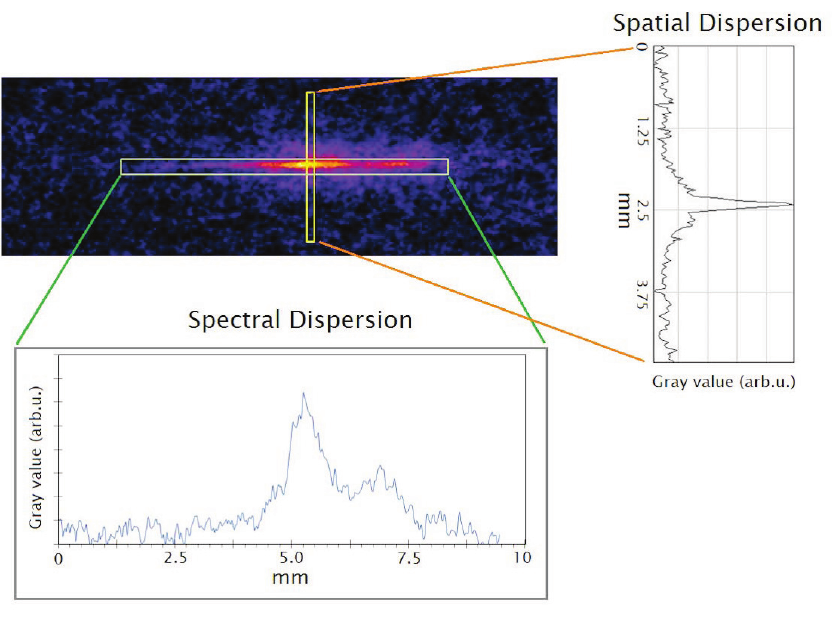} \\[2pt]
  \caption{(Color online) Example of spatially and spectrally resolved K-$\alpha$ emission from the spherical quartz crystal spectrometer. The foil in this case was 10 $\mu$m thick.}\label{spherical}
\end{figure}

Turning to the spatial extent of the K-$\alpha$ source, we see in figure \ref{spherical} a typical spectrum from the rear of a foil, taken with the spherical quartz crystal. Starting with spatial line-outs, we see that there is a central feature sitting on top of a 'pedestal'. This spatial pedestal is a real effect that was seen to disappear when we used a simple wire target in the characterisation of image plate resolution. The width of both features increases with the thickness of the foil as we might expect for a diverging beam of electrons. The fact that the absorption mean free path for the K-$\alpha$ photons is approximately 20 $\mu$m complicates the analysis a little but since we use foils up to 100 $\mu$m thick, the data does indicate that the increasing width of the central feature is indeed a result of a divergent electron beam penetrating the foil. 

\begin{figure}[htbp]
\centering
  \includegraphics[width=7.5cm]{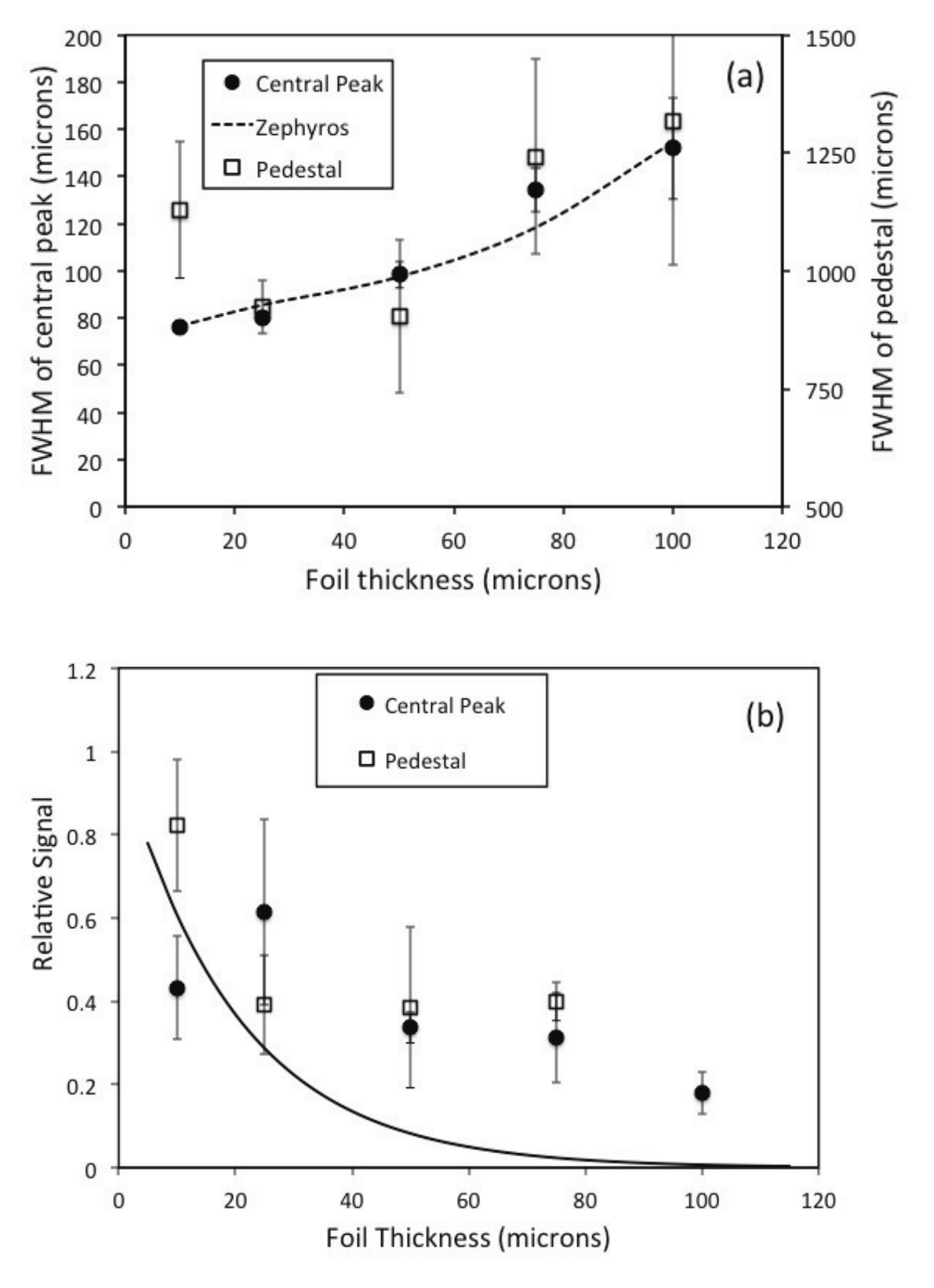} \\[2pt]
  \caption{(a) K-$\alpha$ source size for both the central component and the 'pedestal' as a function of foil thickness. Values are corrected for the resolution. The dashed line is from Zephyros simulations as discussed in the text.  (b) Relative integrated signal from central peak and pedestal features. The solid line is an exponential decay with 20 $\mu$m decay length corresponding to the mean free path of K-$\alpha$ photons in Ti.}\label{divergence}
\end{figure}

The spatial pedestal is seen in both thick and thin foils and so is not thought to be primarily an artefact of lateral spreading of electrons at the laser-irradiated surface. Spreading due to surface currents has been shown to be asymmetric depending on the angle of incidence \citep{li2,gray}. Our spatial resolution is in the direction perpendicular to the incident plane and so would not see this asymmetry. However we can rule out this origin by considering figure \ref{divergence} (a) where we see the size of the measured features as a function of foil thickness. The values are averages of several shots in each case and have been adjusted for instrument resolution by assuming the originals are broadened by convolving in quadrature a $70 \mu$m FWHM Gaussian, c.f. section IIA. By fitting to the data for thicker foils, from 50-100 $\mu$m, we see that the central K-$\alpha$ feature indicates a half-angle divergence of $25-30^{\circ}$. Also shown is the predicted K-$\alpha$ spatial width determined from simulations with the Zephyros code, as discussed below. In figure \ref{divergence}(b) we see the normalised and integrated signal coming from both features as a function of foil thickness. Since we only have one dimensional spatial resolution, we estimate the relative signals in the pedestal and central peak by taking the peak values and multiplying by the square of the experimental FWHM. We then normalise to the energy of the shot. The solid line is an exponential with a 20$\mu$m decay length that illustrates how the signal should decay if it came only from the laser irradiated surface of the foil. We see that, in both cases, the integrated signal decays only slowly with distance. This is due to the penetration of the electrons into the foil and supports the assertion that the pedestal is not simply a result of lateral spreading on the front surface of the target. We discuss its origin further below. 

\begin{figure}[htbp]
\centering
  \includegraphics[width=7.5cm]{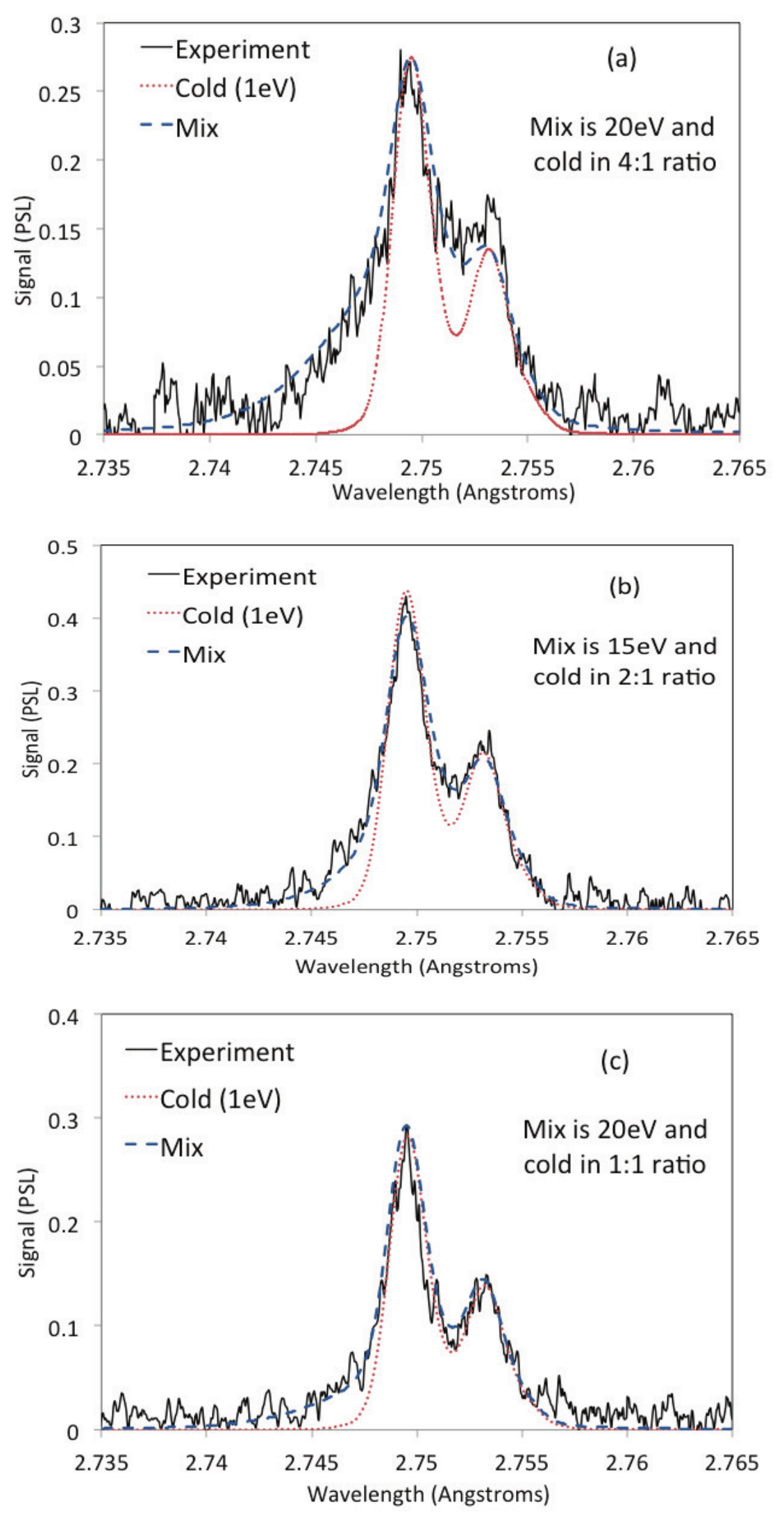} \\[2pt]
  \caption{(Color online) (a) Spectral line-out from 10 $\mu$m foil compared to simulation from the SCRAM code as described in the text. (b) similar comparison for 25 $\mu$m foil, where we see a much smaller influence of higher temperature regions. (c) For the 50 $\mu$m case higher temperature contribution is less important. For thicker targets we found that the higher temperature contribution was not discernible, although the data is relatively noisy due to lower signal levels.}\label{lineouts}
\end{figure}

Regarding the K-$\alpha$ spectrum, figure \ref{lineouts} shows spectral line-outs from the centre of  images for 10, 25 and 50 $\mu$m foils. The strong doublet feature typical of cold K-$\alpha$ (ratio of 2:1 with separation of $\sim 3.7$m\AA) seems to be present for all thicknesses, indicating that colder material is contributing in all cases.  This may be due to the fact that, although we have spatial resolution in one dimension, in the perpendicular direction the spectrum is integrated across the K-$\alpha$ feature and thus the spectra have contributions from emission away from the centre, including the pedestal region.

We have compared the experimental spectra to artificial spectra generated with the SCRAM code \citep{hansen1,hansen2} for a series of solid target temperatures ranging from 1-100eV.  We expect there to be a gradient of temperature and so do not expect a perfect fit to one temperature, however, we can get an indication of the background temperature adopting a two temperature model.  In figure \ref{lineouts}(a), for a 10$\mu$m foil we have fitted to an experimental spectrum by assuming the emitting material is 1 part solid at 1eV and 4 parts solid at 20eV. The doublet ratio is not quite right but the short wavelength shoulder fits quite well.  Also in figure \ref{lineouts} we can see similar plots for 25 $\mu$m and 50 $\mu$m foils. In these cases, the short-wavelength shoulder is less obvious and the dip between the doublet is more pronounced. For the 25$\mu$m foil case, we found a reasonable two temperature fit by mixing solid at 15eV with cold solid in a ratio of 2:1 and for the 50 $\mu$m foil case, we find that mixing 20eV solid and cold solid in a one to one ratio allows a reasonable fit. In all cases, these are indications, we expect there to be gradients and temporal averaging that means  obtaining a unique combination of temperatures would be impractical. Nevertheless, it is clear that increasing foil thickness means we sample a greater proportion of colder material on average, as expected from the reflux data.

\begin{figure*}[htbp]
\centering
  \includegraphics[width=16cm]{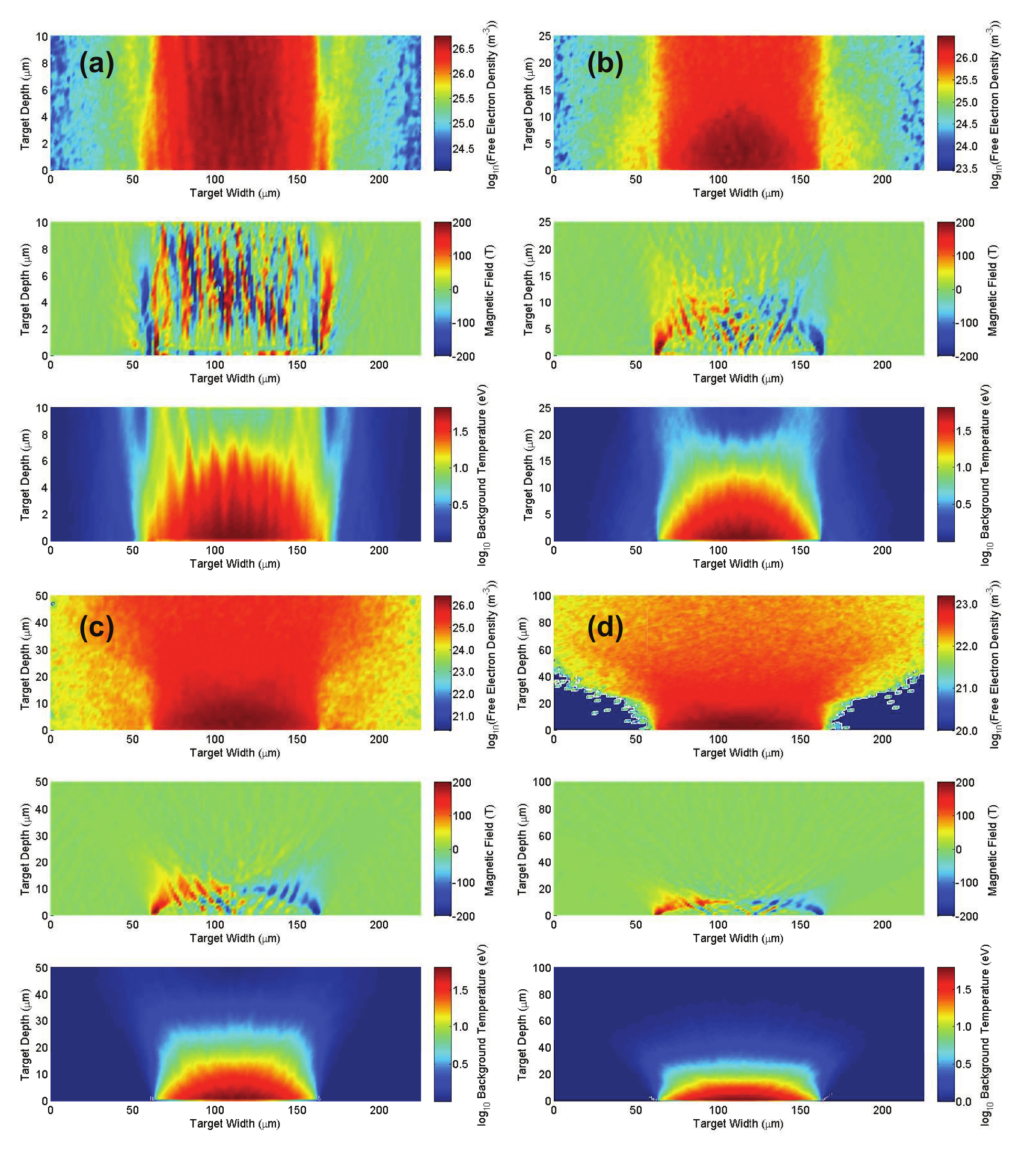} \\[2pt]
  \caption{(Color online) Zephyros simulations of fast electron density for foil thicknesses from 10-100 $\mu$m.  The vertical axis, about which the simulation is symmetric is the normal to the target and the laser is incident from the bottom. For each case, we show magnetic field perpendicular to the target normal direction, the fast electron density and the background temperature.  The absorption is set to $30\%$ and the input divergence half-angle of the fast electrons is set to $60^{\circ}$. The time of the snapshots is at 0.8ps, just at the end of the pulse.}\label{zephyros}
\end{figure*}

\section{Zephyros Simulations}
Figure \ref{zephyros}, shows results from simulations with the 3-D macro-particle hybrid code, Zephyros \citep{zephyros1,zephyros2} where the foil thickness is set to 10, 25, 50 and 100$\mu$m. The images presented are snapshots taken at 0.8ps after the start of the simulation, which corresponds to the end of the input laser pulse. This simulation does not model the laser-plasma interaction but inputs a beam of electrons at a given temperature that depends on the laser intensity, with an initial divergence. Based on the efficiency analysis above, we have assumed $30\%$ absorption of the laser energy into fast electrons. In previous work, it has been noted that the K-$\alpha$ spot size is always bigger than the optical spot size \citep{eder}. This is believed to be due to ejection of fast electrons from the focal region followed by re-entry to the foil away from the laser focus. In order to account for this, the injection region is defined to have a larger radius. The minimum source size measured by us was approximately 70 $\mu$m after accounting for resolution. This is of the order seen by others \citep{eder,park2}.  We assume here that this size results from lateral spread on the front illuminated side as electrons are initially accelerated down the density gradient and not from refluxing. The initial injection of the electrons is made in a uniform cone with $60^{\circ}$ half-angle divergence. This divergence is arrived at by running a series of simulations with different values and comparing how well simulated growth of the K-$\alpha$ feature compares to experiment, as seen in figure \ref{divergence}. In carrying out this comparison, we have assumed that the K-$\alpha$ emission is proportional to the fast electron density. We then post process the Zephyros data to account for the escape depth of the K-$\alpha$ photons and calculate the expected width of the emission using the whole foil depth and integrating over output time-steps (every 0.2ps up to 2.4ps). This approximation is equivalent to assuming weak dependence on ionisation cross-section with energy or that the energy spread of fast electrons is quite uniform spatially. These calculations are carried out for each simulated foil thickness individually to allow for appropriate refluxing to contribute to the results at each thickness.

Some things are immediately obvious from the simulation in figure \ref{zephyros}. Firstly, we see that we do indeed expect refluxing to be far less significant in the case of 100 $\mu$m foils since fewer of the fast electrons reach the rear than for thinner foils. This explains our data with and without an epoxy layer. We also see that the waist of the electron beam stays relatively constant for the initial 20 $\mu$m before diverging and this is reflected in the similar  K-$\alpha$ spot size for 10 and 25 $\mu$m foils, as seen in simulation and experiment. This is likely to be due to the strong magnetic field generated, which can also be seen in fig \ref{zephyros} and reaches over 300T, meaning that a typical gyroradius is of the order a few microns in the foil. An important thing to notice is that for the 10 $\mu$m case, the structure of the magnetic field looks different. In the thicker foils, there is a clearer large scale azimuthal structure that we believe collimates the electron beam. For the 10 $\mu$m case there seems to be strong evidence of filaments forming and a lack of a larger scale azimuthal field structure. Simulations run for 10$\mu$m foils with refluxing switched off show the collimating field structure reappear, confirming that the refluxing is responsible for the difference with thin foils. This is a phenomenon that has been discussed by Yuan {\it et al} \citep{yuan} who discussed the role of refluxing in inhibiting the growth of a collimating magnetic field in thin foils. The fast electron beam still seems to be contained however, probably due to the localised beam filaments creating strong B-fields and the small distance of the rear of the foil. 

We note that the input divergence used in Zephyros, to reproduce the experimental spatial width of the K-$\alpha$ feature, seen in figure \ref{divergence} is about twice the divergence seen experimentally. This is an indication of a strong collimating effect on the fast electrons due to the collective B-field. The effect of self generated B-fields in collimating a beam of electrons has also been discussed in the paper by Yuan {\it et al}\citep{yuan}. In that work, collimation was inferred using experimental results on proton acceleration from variable thicknesses of target foil. The collimation observed in the simulations of Yuan {\it et al}  were present over a much longer distance than seen in our simulations; but this is expected since their intensities were above $10^{20} Wcm^{2}$ and the self generated B-fields were an order of magnitude greater than in the present work. 

Returning to the issue of the K-$\alpha$ emission pedestal, discussed above, we can note that the input divergence used is broadly similar to the divergence seen for the pedestal feature. This leads to the suggestion that the pedestal is formed from fast electrons that have been injected into the foil but escaped the collimating influence of the magnetic field. The Zephyros simulations for thicker foils, at early time do indicate a much higher divergence of fast electrons (greater than 90$^{\circ}$ full angle at 0.2ps) with the magnetic field $<$100T. The Zephyros simulation has a flat top pulse profile and so even at early time we reach the peak irradiance. It is possible that since, experimentally, the pulse will rise in a quasi-Gaussian profile, that significant fast electrons can be injected before the magnetic field is strong enough to have a collimating effect and these fast electrons at early time cause the pedestal feature. This is difficult to investigate conclusively with more simulation, partly due to the large size of the pedestal region seen experimentally but also due to the fact that the simulation does not deal with the details of electron injection from the pre-formed plasma and this is likely to be a key factor. 

From figure \ref{zephyros} we can see that the background bulk solid temperature peaks at around 60eV with strong gradients. For thinner foils, we expect that there will be strong gradients and higher temperature contributions to the K-$\alpha$ spectrum as was seen in figure \ref{lineouts}. However, it is evident that for thicker foils of 50 $\mu$m or more, we expect a significantly reduced contribution from the hot region due to self- absorption as was seen experimentally. Comparison of experiment with SCRAM modelling does not seem to require temperatures as high as 60eV. This may be because of an overestimate of the temperature by Zephyros, but the contribution of expected strong gradients in temperature and integration across the focal spot may also serve to obscure higher temperature contributions. In summary, it does seem that heating to 10 s of eV is seen as expected and that the domination of emission from colder bulk temperature does increase with foil thickness, consistent with the data on refluxing.

\section{Discussion and Summary} 
In this work, we have noted several features of the dynamics of fast electrons in intense laser-foil interactions. We have observed the important contribution of refluxing to both the K-$\alpha$ and hard x-ray signals. The effect on K-$\alpha$ has been noted before \citep{nersisyan, neumayer}, but the more muted effect on hard x-rays has been less noted. We have also, by comparison to modelling seen that the number of 'bounces' is apparently limited, although, given the relatively low electron energies, this is consistent with previous work. We have observed that not only is the hard x-ray yield reduced by the epoxy layer but there is a clear effect on the fast electron temperature measurement for thinner foils. It was shown that this is unlikely to be caused by loss of faster electrons into the low Z epoxy. We also considered the possibility that early refluxing fast electrons could reheat the pre-formed plasma before the peak of the main pulse, thus altering the fast electron temperature generated later in the pulse. However, simple analysis  based on stopping power of fast electrons suggested the effect would again be insufficient. However, we can also estimate that for a sufficient heating of the preformed plasma below critical density only a fraction of a percent of the available refluxed fast electron energy would need to be coupled in at  below critical density and some other mechanism for this to occur may be possible. Further work seems needed on this issue, perhaps in terms of a two beam instability caused by interaction of refluxing electrons with electrons injected in the opposite direction. At present we are not equipped to run such simulations.

We have seen that the K- $\alpha$ source size is affected by a diverging electron beam and that this is reproduced in broad terms by simulations using the Zephyros code.  The initial divergence used in Zephyros is about double that seen experimentally. This is taken to indicate that there is collimation of the fast electrons via the self generated magnetic field and indeed, figure \ref{zephyros} shows both a restricted divergence of the electrons and strong magnetic field up to about 30$\mu$m into the foil. The observation of a pedestal in Zephyros was not possible, partly due to size limitations in the simulations. The origin of the pedestal is still not entirely clear but, as discussed, it may be due to electrons injected early in the pulse before a strong collimating field has been generated.

There is still scope for improvements to such measurements.  For example, the one dimensional spatial resolution of the spherical crystal allows us to see the spectrum as well but has the disadvantage of convolving spectral information across the source and better spatial resolution could be achieved. With better spatial resolution, an Abel inversion technique may be useful in unfolding the spectral and spatial contributions e.g.  \citep{cristoforetti}. Recent work \citep{chatterjee} has pointed the way to measuring  the large magnetic fields generated as a way of mapping the filamentation of the electron beam that would be useful in comparing thin and thick foils regarding the role of refluxing in the magnetic field structure.

\section{Acknowledgements} This work was supported by the UK Engineering and Physical Sciences Research Council (grant Nos. EP/C003586/1, EP/G007462/01 and EP/I029206/1)

\end{document}